\documentstyle[preprint,revtex,eqsecnum]{aps}
\begin{document}
\draft
\baselineskip = 0.9\baselineskip
\begin{title}
{Phase transitions in anisotropic superconducting \\
and magnetic 
systems with vector order parameter: \\
three-loop renormalization group analysis}
\end{title}
\author{S. A. Antonenko  and A. I. Sokolov}
\begin{instit}
Department of Physical Electronics, Electrotechnical University, \\
Prof.Popov Str. 5, St.-Petersburg, 197376, Russia
\end{instit}
\begin{abstract}
The critical behavior of the model with $N$--vector complex order 
parameter and three quartic coupling constants which describes 
phase transitions 
in unconventional superconductors, helical magnets, stacked 
triangular 
antiferromagnets, superfluid helium--3 and zero--temperature 
transitions in fully frustrated Josephson--junction 
arrays is studied within the field--theoretical 
renormalization group (RG)
approach in three dimensions. To obtain qualitatively 
and quantitatively
correct results perturbative expansions for $\beta$--functions 
and critical
exponents are calculated up to three--loop order and resummed 
by means of
the generalized Pad$\acute {\rm e}$--Borel procedure. Fixed point 
coordinates, critical exponents values, RG flows {\it etc.} are found
for physically interesting cases $N = 2$ and $N = 3$. Critical
(marginal) values of $N$ at which the topology of the flow diagram
changes are determined as well. It is argued, on the base of several
independent criterions, that the accuracy of the numerical results
obtained is about 0.01, an order of magnitude better than that given 
by resummed two--loop RG expansions.

In most cases the systems mentioned are shown to undergo 
fluctuation--driven first--order phase transitions. Continuous
transitions are allowed in hexagonal $d$--wave superconductors,
in planar helical magnets (into sinusoidal linearly--polarized phase),
and in triangular antiferromagnets (into simple unfrustrated ordered
states) with critical exponents $\gamma = 1.336$, $\nu = 0.677$, 
$\alpha = -0.030$, $\beta = 0.347$, $\eta = 0.026$ 
which are hardly believed to be experimentally 
distinguishable from those of the $3D$ $XY$
model. The chiral fixed point of RG equations is found to really
exist and possess some domain of attraction provided $N \ge 4$. So,
magnets with Heisenberg ($N = 3$) and $XY$--like ($N = 2$) spins
would not demonstrate chiral critical behavior with unusual values
of critical exponents; they can approach the chiral state only via
first--order phase transitions.
\end{abstract}
\narrowtext

\section{Introduction}
\label{sec:1}
The renormalization group (RG) approach in three dimensions 
proved to be
very efficient when used to study the critical behavior of simple
$O(n)$--symmetric models. Calculations of higher--order RG expansions
for field--theoretical $\beta$--functions and critical exponents 
combined
with proper resummation of the series obtained resulted in the 
estimates
of critical exponent values which nowadays are referred to as most
accurate (canonical) numbers \cite{1,2}. This approach enabled one to
give quantitatively correct description of critical behavior of more
complex systems possessing two quartic coupling constants in their
Landau--Wilson Hamiltonians. We mean the impure Ising model 
\cite{3,4,5},
the cubic model \cite{4,6}, and the $mn$--model \cite{7}. 
Moreover, the
method turned out to be powerful enough even in two dimensions as was
shown by comparison of the approximate results obtained on the base of
four--loop RG expansions resummed by means of 
Pad$\acute {\rm e}$--Borel--like technique
with their exact counterparts known for (exactly solvable) $2D$ 
Ising and impure Ising models \cite{1,2,5}.

On the other hand, there are numerous models with more than two 
quartic
coupling constants which describe phase transitions in a variety of
systems.  
Such
models, however, being extensively studied in the frame of RG
approach were actually treated only within the lowest--order 
(one-- and
two--loop) approximations which are known to lead to rather crude
quantitative and, sometimes, to contradictory qualitative results.

The aim of this paper is to study the static critical behavior of the
three--dimensional model with three quartic coupling constants on 
the base
of three--loop RG series resummed in the way which provides proper 
physical
predictions and accurate numerical estimates. As far as we know, 
this is
the first attempt to get reliable, numerically correct results for a 
complicated $3D$ field--theoretical model from higher--order RG 
expansions \cite{7a}. The Landau--Wilson Hamiltonian 
of the model is as follows:
\FL
\begin{eqnarray}
H = {1 \over 2} 
\int d^3x \Bigl[ m_0^2 \varphi_{\alpha} \varphi_{\alpha}^*
 + \nabla \varphi_{\alpha} \nabla \varphi_{\alpha}^*  
+ {u_0 \over 2} \varphi_{\alpha} \varphi_{\alpha}^* \varphi_{\beta} 
\varphi_{\beta}^* \nonumber \\ 
+\ {v_0 \over 2} \varphi_{\alpha} \varphi_{\alpha} \varphi_{\alpha}^* 
\varphi_{\alpha}^* 
+ {w_0 \over 2} \varphi_{\alpha} \varphi_{\alpha} \varphi_{\beta}^* 
\varphi_{\beta}^* \Bigr] \ \ , \ \ \ \label{eq:1.1}
\end{eqnarray}
where $\varphi_{\alpha}$ is a complex vector order parameter field, 
$\alpha, \beta = 1, 2,$ \ldots, $N$, a bare mass squared $m_0^2$ being
proportional to the deviation from the mean--field transition point 
(line).

This model describes critical phenomena 
in a plenty of substances. Their list includes tetragonal,
hexagonal and cubic superconductors with $d$-- or $p$--wave pairing 
\cite{8}
as well as superconductors with two --- $s$ and $d$ --- order 
parameters \cite{9}, fully frustrated
Josephson--junction arrays (FFJJA) at zero temperature \cite{10}, 
stacked
triangular antiferromagnets (STA) \cite{11,12}, helical magnets (HM)
and magnets with sinusoidal spin structures \cite{12,13,14,15}, 
the A phase of superfluid ${}^3$He \cite{16,17}.
The model Eq.~(\ref{eq:1.1}) is related also to the critical 
thermodynamics of type--II superconductors with short 
coherence length near the upper critical magnetic field \cite{18}.

All the systems mentioned were extensively studied during the 
last decade
and rich theoretical information about their critical behavior 
has been obtained
both analytically and numerically. 
Unfortunately, the
major part of these data turns out to be contradictory or 
inconclusive. To illustrate this point we overview, 
in brief, what was predicted for
FFJJA, STA and HM by different people within different approaches.

A superconductor--insulator transition in FFJJA at zero temperature 
produced by competition of Josephson and charging effects in the 
presence of quantum fluctuations is described by three--dimensional 
model Eq.(\ref{eq:1.1}) with $N = 2$ and $v_0 = 0$ or
$w_0 = 0$. Starting from $4 - \epsilon$ dimensions, such a transition 
was shown to be, within the lowest order in
$\epsilon$, discontinuous \cite{22}, while $(2 + \epsilon)$--expansion
did not enable one to resolve whether it should be first--order or 
continuous \cite{22}. On the other hand, this transition was referred 
to as second--order one on the base of 
analysis valid to the leading order in $1 / N$ \cite{10}.

For $N = 2$ and $N = 3$, the Hamiltonian (\ref{eq:1.1}) 
governs the critical 
behavior of STA such as VCl$_2$, VBr$_2$, CsMnBr$_3$, CsVCl$_3$, 
and of HM (Ho, Dy, Tb, $\beta$--MnO$_2$, MnAu$_2$). 
In the case of Heisenberg spins, RG calculations in
$4 - \epsilon$ dimensions and $1 / N$--expansion result in a 
first--order
phase transition \cite{13,14,15,27}, although 
$\epsilon$--expansion predictions
were believed also as favoring continuous one \cite{12}. Monte--Carlo
simulations in $3D$ seem to provide an evidence of continuous phase 
transition \cite{11,28}. RG analysis of corresponding 
$(2 + \epsilon)$--dimensional model proposes that systems mentioned 
should
undergo, in three dimensions, a first--order transition or a 
second--order
one with either $O(4)$ (not $O(6)$) critical or tricritical 
mean--field exponents \cite{29}.

Obviously, the situation needs to be cleared up. Since the 
problem does not allow exact solution, 
in order to obtain reliable theoretical information
one has to employ approximate methods with controlled or, 
at least, known
level of accuracy. Regular RG perturbation theory in $3D$ 
subject to the
application of Pad$\acute {\rm e}$--Borel--like resummation technique 
will be shown to play
a role of such a method. 

The paper is organized as follows. In Sec.~\ref{sec:2} 
the Hamiltonians describing the 
systems mentioned above are considered and related to the
Hamiltonian (\ref{eq:1.1}). In Sec.~\ref{sec:3} the renormalization 
scheme is formulated
and three--loop RG expansions for $\beta$--functions and critical 
exponents are presented. Various resummation 
techniques based on the Borel transformation 
and applicable to divergent (asymptotic) power series of
several independent variables are considered and criterions 
for the choise of the best one are established. 
The specific symmetries of the model (\ref{eq:1.1}) with $N = 2$ 
are discussed in detail. They 
relate coordinates of different fixed 
points of RG equations to each other being a sensitive
indicator of the quality of the approximation employed. 
All the numerical results obtained are presented 
in Sec.~\ref{sec:4}: coordinates of the fixed points,
critical exponent values, critical (marginal) 
order parameter
dimensionalities $N_c$ at which the topology of flow diagrams changes, 
{\it etc}. RG flows are also shown in the planes where 
stable, within these planes, fixed points exist. 
In Sec.~\ref{sec:5} the results obtained are applied to  
superconducting, superfluid and magnetically--order systems 
and certain 
theoretical predictions are made. Particular attention is 
paid to what is known as chiral critical behavior and 
its relevance to real HM and STA with Heisenberg or $XY$--like spins. 
Sec.~\ref{sec:6} containes a summary of the results obtained. 
Details of Pad$\acute {\rm e}$--Borel resummation procedure
are described in Appendix.

\section{Relevant substances and structures}
\label{sec:2}

In this section we discuss physical systems undergoing phase 
transitions which are described by the Hamiltonian (\ref{eq:1.1}).

\subsection{Unconventional superconductors}

These materials should be mentioned first since Eq.~(\ref{eq:1.1}) 
is actually an
obvious generalization of appropriate Ginzburg--Landau form (see, e.g.
Ref.~\cite{8}) with $\varphi_{\alpha}$ being a superconducting order 
parameter. For
$N = 2$ the Hamiltonian under consideration governs a static critical
behavior of tetragonal and hexagonal ($v_0 = 0$) superconductors with
$d$--wave pairing, while the case $N = 3$ corresponds to 
cubic $p$--wave
materials. Heavy--fermion compounds such as UPt$_3$, UBe$_{13}$,
and others are thought to belong to 
this class
of superconductors \cite{31,33}. Phase transitions in 
thorium--doped UBe$_{13}$
are well described by the phenomenological model with two coexisting,
$s$-- and $d$--wave, order parameters \cite{9,35} which, in some 
limit, is reduced to that given by Eq.~(\ref{eq:1.1}). 
Moreover, since there are numerous
experimental facts \cite{40,40a} and theoretical 
predictions \cite{41,43a} favoring non--trivial 
pairing modes in high--$T_c$ superconductors, 
the Hamiltonian (\ref{eq:1.1}) may be also 
relevant to the critical behavior of these new materials \cite{43b}.

It is worthy to note that the width of critical region is large 
enough in high--$T_c$ superconductors 
(see Refs.~\cite{44,45} for an overview and numerical
estimates) and superconducting fluctuations proved to be clearly 
seen in their thermodynamics near $T_c$ \cite{46,47,48,49}. 
Wide fluctuation regions are also expected to exist 
in heavy--fermion compounds \cite{51,52}. That is why
the critical behavior of the model (\ref{eq:1.1}) is 
extensively studied within the
context of superconductivity \cite{53,54,55}. On the other hand, the 
Hamiltonian (\ref{eq:1.1}) has only one, isotropic gradient 
invariant, i.e. it ignores a crystallographic anisotropy 
of the order parameter correlation
function which may play an appreciable role in the critical region. 
So, the applicability of the Eq.~(\ref{eq:1.1}) to real 
unconventional superconductors
is somewhat limited. The influence of anisotropic gradient terms on
thermodynamics of these materials in the region of weak, Gaussian
fluctuation was studied in Ref.~\cite{56}.

\subsection{Fully frustrated Josephson--junction arrays}

Main features of JJA behavior are known to be 
described by the following Hamiltonian \cite{57}:
\FL
\begin{equation}
H = - {E_c \over 2} \sum_i {\partial \overwithdelims () 
{\partial \theta_i}}^2 - E_J \sum_{<ij>} \cos (\theta_i - \theta_j - 
A_{ij}) \ \ , \label{eq:2.1}
\end{equation}
where $\theta_i$ is a phase of superconducting order parameter in 
$i$--th island,
\begin{equation}
A_{ij} = {{2 \pi} \over \Phi_0} \int\limits_i^j {\bf A} 
d{\bf \ell} \ \ ,
\label{eq:2.2}
\end{equation}
$\bf A$ being a vector potential of external magnetic field, 
and $\Phi_0$ is a quantum of flux. Here $E_c$ plays 
a role of charging energy which is responsible for the Coulomb 
blockade and quantum dynamics while the Josephson coupling 
$E_J$ favors establishing of the global phase coherence
and overall superconductivity in the system. At zero temperature
superconductor--to--insulator transition occurs when the ratio 
$E_c / E_J$ exceeds a critical value. Since quantum 
fluctuations are essential in the case considered, 
the effective dimensionality of the system turns out to be
equal to three: $D = 2 + 1$ (see, e.g. Ref.~\cite{10}).

If the external magnetic field $\bf B$ is uniform JJA is regulary 
frustrated with the frustration parameter $f = {(B a_0) / \Phi_0}$, 
$a_0$
being an area of a plaquette. We shall consider JJA with square and
triangular lattices in a magnetic field corresponding to 
$f = {1 \over 2}$ which are usually referred to as fully frustrated 
ones. To study their critical behavior 
a proper Hubbard--Stratonovich transformation \cite{10,60a}
may be applied to the model (\ref{eq:2.1}) resulting in the 
Landau--Wilson Hamiltonian with quartic terms which are, 
in notations of Ref.~\cite{10}, as follows:
\FL
\begin{equation}
u \bigl( {| \psi_1 |}^2 + {| \psi_2 |}^2 \bigr)^2 - v_1 
{| \psi_1 |}^2
{| \psi_2 |}^2 + v_2 {\rm Re}(\psi_1^* \psi_2)^2 \ \ , \label{eq:2.3}
\end{equation}
where $\psi_1$ and $\psi_2$ are complex scalar fields. In the case of
square lattice $u > 0$, $v_1 = v_2 >0$, while for the triangular FFJJA
$u > 0$, $v_1 < 0$ and $v_2 = 0$. It is easy to see that 
Eq.~(\ref{eq:2.3}) is actually identical to the quartic part 
of the Hamiltonian (\ref{eq:1.1})
for $N = 2$ provided coupling constants are related to those standing
in Eq.~(\ref{eq:1.1}) by
\FL
\begin{equation}
u = u_0 + v_0 + w_0 \ , \quad v_1 = 2(v_0 + w_0) \ , 
\quad v_2 = 2w_0 \ \ . 
\label{eq:2.4}
\end{equation}
Domains $v_0 = 0$, $w_0 > 0$ and $v_0 < 0$, $w_0 = 0$ correspond 
to the square and triangular FFJJA respectively. 
The Hamiltonian (\ref{eq:1.1})
governs also the critical behavior of triangular JJA with 
$f = {1 \over 4}$ since it is known to belong to the same 
universality class as FFJJA with square lattice \cite{60a}.

\subsection{Stacked triangular antiferromagnets}

Triangular antiferromagnets which we shall deal with possess lattices
consisting of triangular antiferromagnetic layers stacked in register
along the orthogonal axis. In the ground state the spin arrangement 
may be thought as formed by three ferromagnetic sublattices 
with $120^{\circ}$ angles between neighboring, 
within the layer, spins (see Refs.~\cite{12,58}
for detail). The microscopic Hamiltonian modelling STA reads 
\cite{12}: 
\FL
\begin{equation}
H = - J \sum_{<ij>} {\bf S}_i {\bf S}_j - J^{\prime} 
\sum_{<ij>^{\prime}}
{\bf S}_i {\bf S}_j \ \ , \quad J < 0 \ \ . \label{eq:2.5}
\end{equation}
The first sum represents antiferromagnetic interactions within 
triangular layers which give rise to frustration. 
The second one describes interlayer coupling, 
the sign of $J^{\prime}$ being unimportant since there is no
frustration along the orthogonal direction. The Hubbard--Stratonovich
transformation followed by the expanding around the instability 
points and
other standart procedures leads to effective Hamiltonian containing
\begin{equation}
u_k ({\bf a}^2 + {\bf b}^2) + v_k \bigl[({\bf a b})^2 -
{\bf a}^2 {\bf b}^2 \bigr] \label{eq:2.6}
\end{equation}
as an interaction term, ${\bf a}$, ${\bf b}$ being real $n$--component
vector fields \cite{12} . If then one put
\begin{equation}
u_k = u_0 + w_0 \ \ , \qquad v_k = 4w_0 \label{eq:2.7}
\end{equation}
Eq.~(\ref{eq:2.6}) will immediately turn into the quartic part of the
Hamiltonian (\ref{eq:1.1}) with $v_0 = 0$ and 
$\varphi_{\alpha} = a_{\alpha} + i b_{\alpha}$. The frustration may be
shown to be relevant only for $w_0 > 0$; the opposite case, 
$w_0 < 0$, corresponds to simple ferromagnetic or antiferromagnetic 
ordering \cite{58}.

\subsection{Helical magnets}

In these magnets spins aling ferromagnetically in a plane and form 
spirals
along the orthogonal axis. Such an ordering may be described by the
microscopic Hamiltonian (\ref{eq:2.5}) provided the first and second
sums are defined in a new manner \cite{12}. Namely, let the first sum
represents nearest--neighbor ferromagnetic interactions, $J > 0$, 
while
the second term in (\ref{eq:2.5}) describes antiferromagnetic, 
$J^{\prime} < 0$, next--nearest--neighbor interactions acting 
along only
one crystallographic axis. Then for ratios $|J^{\prime}| / J$ 
exceeding a critical value spins will be helically arranged 
along this axis. All
the machinery mentioned above gives in this case just the same
Landau--Wilson Hamiltonian as for STA. The helical ordering, however,
is realized only if $v_k > 0$ ($w_0 > 0$) \cite{12}.
For $v_k < 0$ a sinusoidal
(linearly--polarized) spin density wave should occur \cite{15}.

\subsection{Superfluid helium--3}

In liquid ${}^3$He fermionic excitations are known to form, 
below $T_c$,
Cooper--like pairs with $S = L = 1$. Since magnetic dipole--dipole
interaction couples orbital and spin angular momenta to each other the
superfluid order parameter possesses a symmetry $O(3) \times U(1)$.
This is precisely the symmetry underlying the Hamiltonian 
(\ref{eq:1.1})
with $N = 3$ and $v_0 = 0$. As was shown in Refs.~\cite{16,17}, 
Eq.~(\ref{eq:1.1}) describes, in fact, the transition of liquid 
helium--3
from normal to superfluid Anderson--Morel phase; coupling constants
$g_0$ and ${\lambda}_0$ entering formulas of Refs.~\cite{16,17} 
are easily
seen to be identical to $u_0$ and $w_0$ respectively.

\section{RG series, resummation and symmetries}
\label{sec:3}

As was already mentioned, the static critical behavior of the model 
Eq.~(\ref{eq:1.1}) has been studied in three dimensions within one--
and two--loop RG approximations \cite{55}. The taking into account of
two--loop contributions to the $\beta$--functions  and critical 
exponents
was found to change drastically  the results of the lowest--order RG
analysis. In particular, it alters the total number of fixed  points
and avoids the degeneracy of the $O(2N)$--symmetric fixed point which
is four--fold degenerate, for $N = 2$, within the parquette 
approximation.
On the other hand, some of the numerical results obtained on the base
of the resummed two--loop RG expansions do not obey some exact 
symmetry relations (see below). In such a situation three--loop 
calculations turn
out to be very desirable.

We calculate the $\beta$--functions for the Hamiltonian 
Eq.~(\ref{eq:1.1}) within a massive theory. 
The renormalized Green function $G_R(p,m)$ and
four--point vertex functions $U_R(p_i,m,u,v,w)$, $V_R(p_i,m,u,v,w)$, 
$W_R(p_i,m,u,v,w)$ are normalized at zero momenta in a conventional 
way:
\begin{eqnarray}
G_R^{-1}(0,m) = m^2 \ \ , \ \ \ \ \ \ \ \ 
\nonumber \\
{{\partial G_R^{-1}(p,m)} \over {\partial p^2}} 
\Big\arrowvert_{p^2 = 0} = 1 \ \ , \ \nonumber \\ 
U_R(0,m,u,v,w) = mu \ \ , \label{eq:3.1} \\
V_R(0,m,u,v,w) = mv \ \ , \nonumber \\
W_R(0,m,u,v,w) = mw \ \ . \nonumber 
\end{eqnarray}
One extra condition is imposed on the $\varphi^2$ insertion:
\begin{equation}
\Gamma_R^{(1,2)}(p,q,m,u,v,w) \Big\arrowvert_{p = q = 0} = 1 \ \ . 
\label{eq:3.2}
\end{equation}
The value of the one--loop vertex graph at zero external momenta 
including the factor $(N + 4)$ is
absorbed in $u$, $v$, $w$ in order to make the coefficient for 
$u^2$ term
in $\beta_u$ equal to unity. The $\beta$--functions obtained are as 
follows: 
\widetext
\begin{mathletters}
\FL
\begin{eqnarray}
\beta_u = u - u^2 - {4 \over {N + 4}} \bigl( uv + uw + w^2 \bigr) + 
{2 \over {27 ( N + 4 )^2}} \biggl[ \bigl( 41\ N + 95 \bigr) u^3 + 
200 u^2v + 200 u^2w \qquad 
\nonumber \\
+ 46 uv^2 + \bigl( 46\ N + 216 \bigr) uw^2 + 92 uvw + 144 vw^2 + 
( 36\ N + 72 ) w^3 \biggr] \qquad \qquad \qquad \qquad \ \ 
\nonumber \\ 
- {1 \over {4 ( N + 4 )^3}} \biggl[ \bigl( 2.69789\ N^2 + 
54.94038\ N + 99.82021 \bigr) u^4 + \bigl( 26.58751\ N 
\qquad \qquad \qquad \quad 
\nonumber \\ 
+ 329.22770 \bigr) \bigl( u^3v + u^3w \bigr) + 
\bigl( 2.48756\ N + 
221.36225 \bigr) \bigl( u^2v^2 + 2 u^2vw \bigr) 
+ \bigl( 2.48756\ N^2 \quad \ 
\nonumber \\
+ 155.55980\ N + 
470.42246 \bigr) u^2w^2 + 50.50080 \bigl( uv^3 
+ 3 uv^2w \bigr) + \bigl( 34.28057\ N \qquad \qquad \quad
\nonumber \\
+ 626.66599 \bigr) uvw^2 + 
\bigl( 8.11011\ N^2 
+ 125.31213\ N + 311.16081 \bigr) uw^3 
+ 110.42034 v^2w^2 \quad
\nonumber \\ 
+ \bigl( 1.95355\ N 
+ 216.93358 \bigr) vw^3 + \bigl( -5.20190\ N^2 - 0.62829\ N 
+ 95.22334 \bigr) w^4 \biggr] \ \ , \qquad \ \ 
\label{eq:3.3a}
\end{eqnarray}
\FL
\begin{eqnarray}
\beta_v = v \Biggl[ 1 - {2 \over {N + 4 }} \bigl( 3u + {5 \over 2}v 
+ 4w \bigr) + {2 \over {27 ( N + 4 )^2 }} 
\biggl[ \bigl( 23\ N + 185 \bigr) u^2 + 
362 uv + 524 uw \qquad \ \ \ \ 
\nonumber \\
+ 136 v^2 + 380 vw + \bigl( 28\ N + 180 \bigr) w^2 \biggr] - 
{1 \over {4 ( N + 4 )^3 }} \biggl[ \bigl( - 2.50221\ N^2 
\qquad \qquad \qquad \qquad \quad 
\nonumber \\
+ 41.85390\ N + 234.66699 \bigr) u^3 + 
\bigl( - 0.01437\ N + 720.91540 \bigr) u^2v 
+ \bigl( 8.98498\ N \qquad \quad \ \ \ 
\nonumber \\
+ 1015.38106 \bigr) u^2w + 
579.33309 uv^2 + 1575.28532 uvw 
+ \bigl( 151.47423\ N \qquad \qquad \qquad \quad \ \ 
\nonumber \\
+ 780.92014 \bigr) uw^2 
+ 157.45847 v^3 + 604.53412 v^2w 
+ \bigl( 6.49576\ N 
+ 753.08966 \bigr) vw^2 \qquad \ \ 
\nonumber \\ 
+ \bigl( - 3.27046\ N^2 + 13.63522\ N 
+ 284.67391 \bigr) w^3 \biggr] \Biggr] \ \ , 
\qquad \qquad \qquad \qquad \qquad \qquad \qquad \ \ \ 
\label{eq:3.3b}
\end{eqnarray}
\FL
\begin{eqnarray}
\beta_w = w \Biggl[ 1 - {2 \over {N + 4}} \bigl( 3u + v + 
{N \over 2} w \bigr) + {2 \over {27 ( N + 4 )^2 }} 
\biggl[ \bigl( 23\ N + 185 \bigr) u^2 + 
200 uv + \bigl( 54\ N \qquad \ \ \ 
\nonumber \\
+ 92 \bigr) uw 
+ 28 v^2 + 56 vw + \bigl( 36 - 8\ N \bigr) w^2 \biggr] - 
{ 1 \over {4 ( N + 4 )^3 }} \biggl[ \bigl( - 2.50221\ N^2 
+ 41.85390\ N \ \ 
\nonumber \\
+ 234.66699 \bigr) u^3 + 
\bigl( - 9.01372\ N + 426.44974 \bigr) u^2v 
+ \bigl( 2.99978\ N^2 
+ 83.14193\ N \qquad \quad \ 
\nonumber \\
+ 230.13930 \bigr) u^2w 
+ 162.71394 uv^2 
+ \bigl( 29.26715\ N + 266.89358 \bigr) uvw 
+ \bigl( 5.75601\ N^2 \quad \ \ 
\nonumber \\
+ 48.11146\ N 
+ 131.38337 \bigr) uw^2 + 25.29977 v^3 + \bigl( 1.15422\ N 
+ 73.59085 \bigr) v^2w \qquad \qquad \ 
\nonumber \\
+ \bigl( 9.52258\ N + 106.38551 \bigr) vw^2 + 
\bigl( - 1.31497\ N^2 + 10.71074\ N 
+ 58.66955 \bigr) w^3 \biggr] \Biggr] . \quad \ \ \ \ 
\label{eq:3.3c}
\end{eqnarray}
\end{mathletters}
\narrowtext

Such series are known to be divergent, at best asymptotic. 
They contain, however, rich and important 
physical information which may be extracted 
provided some procedure making them convergent is applied. The Borel 
transformation usually plays a role of this procedure. Here we are 
dealing with expansions of quantities depending 
on three variables $u$, $v$, and 
$w$. So, Borel transformation should be taken in the generalized form: 

\FL
\begin{equation}
f( u, v, w ) = \sum_{ijk} c_{ijk} u^i v^j w^k = \int\limits_0^\infty 
e^{-t} F( ut, vt, wt ) dt \ \ , 
\label{eq:3.4}
\end{equation}
where the Borel transform expansion is as follows:
\begin{equation}
F( x, y, z ) = \sum_{ijk} {c_{ijk} \over {( i + j + k ) !}} 
x^i y^j z^k 
\label{eq:3.5}
\end{equation}

To calculate the integral entering Eq.~(\ref{eq:3.4}) one should 
perform an analytical continuation of the Borel--transformed 
expansion. Although
there are several different ways to do it, only two approaches proved
to be efficient in the case of multi--variable RG series \cite{4,5}. 
The first one exploits the so called resolvent series \cite{61}: 
\FL
\begin{equation}
\tilde F ( x, y, z, \lambda ) = \sum_{n = 0}^{\infty} \lambda^n 
\sum_{l = 0}^n \sum_{m = 0}^{n - l} {c_{l, m, n - l - m} x^l y^m 
z^{n - l - m} \over {n !}}
\label{eq:3.6}
\end{equation}
which is actually a series in powers of $\lambda$ with coefficients 
$A_n$ being uniform polynomials of $n$--th order in $u$, $v$, 
and $w$. Pad$\acute {\rm e}$ approximants in 
$\lambda$ $[ L / M ]$ is then used 
and the sum of the series is given by 
\begin{equation}
F ( x, y, z ) = [ L / M ] \Big \arrowvert_{\lambda = 1} 
\label{eq:3.7}
\end{equation}
(see Appendix for detail).
This approximation scheme possesses the remarkable property: for 
$y = z = 0$ ( or $x = z = 0$, or $x = y = 0$ ) expression 
(\ref{eq:3.7}) turns into conventional single--variable 
Pad$\acute {\rm e}$ approximants. 
Hence, all the results obtained for simpler, say, 
$O( n )$--symmetric models hold good within this approach. 

Another way of the analytic continuation is realized through 
the construction 
of the Canterbury approximants invented by Chisholm \cite{62}: 
\FL
\begin{equation}
[ K, L, M / R, P, Q ] = {\sum\limits_{k = 0}^K \sum\limits_{l = 0}^L 
\sum\limits_{m = 0}^M A_{klm} x^k y^l z^m \over {\sum\limits_{r = 0}^R 
\sum\limits_{p = 0}^P \sum\limits_{q = 0}^Q B_{rpq} x^r y^p z^q}} 
\ \ . 
\label{eq:3.8}
\end{equation}
It was found to be rather effective when applied to the impure 
Ising model 
\cite{3,4}, the cubic model \cite{4,6}, and the $mn$--model \cite{7}.

To determine which approximation scheme is the most adequate to our 
problem 
certain criterious should be formulated. We adopt the following ones:

i) the resummation technique chosen should not lead to unphysical 
results, 

ii) new results should be consistent with the most accurate numerical 
estimates for $O( n )$--symmetric and other simple models known up 
today, 

iii) new results should be self--consistent, i.e. numerical values 
of any 
critical exponent calculated by means of the resummation of different 
expansions, say, expansions for $\gamma$ and $\gamma^{-1}$, should be 
identical (as close as possible), 

iv) all (known) symmetries of the problem should be preserved by the 
approximation scheme employed.

The last criterion is of prime importance in the case considered. 
The point 
is that the model Eq.~(\ref{eq:1.1}) for $N = 2$ possesses specific 
symmetry properties. Indeed, if the field $\varphi_{\alpha}$ 
undergoes the transformation 
\begin{equation}
\varphi_1 \to \varphi_1 \ \ , \quad \varphi_2 \to i \varphi_2 \ \ ,
\label{eq:3.9}
\end{equation}
quartic coupling constants are also transformed:
\begin{equation}
u \to u \ \ , \quad v \to v + 2w \ \ , \quad w \to -w \ \ , 
\label{eq:3.10}
\end{equation}
but the structure of the Hamiltonian itself remains unchanged 
\cite{55}.
Just the same situation takes place in the case of another field 
transformation \cite{10,12}: 
\begin{equation}
\varphi_1 \to {{\varphi_1 + i \varphi_2} \over {\sqrt 2}} \ \ , \quad 
\varphi_2 \to {{i \varphi_1 + \varphi_2} \over {\sqrt 2}} \ \ , 
\label{eq:3.11}
\end{equation}
which does not affect the Hamiltonian structure resulting only in the 
following replacement of $u$, $v$, and $w$:
\FL
\begin{equation}
u \to u + v + 2w \ \ , \quad v \to -2w \ \ , \quad w 
\to -{v \over 2} \ \ .
\label{eq:3.12}
\end{equation}

It is well known that RG functions of the problem are completely 
determined by the structure of the Hamiltonian: they do not depend on 
$u_0$, $v_0$, and $w_0$ which play a role of initial values of 
effective 
coupling constants when the RG flow of $u$, $v$, and $w$ is searched. 
Hence, RG equations should be invariant with respect to any 
transformation conserving the structure of the 
Hamiltonian \cite{63}; Eqs.~(\ref{eq:3.10}) 
and (\ref{eq:3.12}), in particular, were shown to be such 
transformations. 

It means that for $N = 2$ \ $\beta_u$, $\beta_v$, and $\beta_w$ 
should obey 
some special symmetry relations which may be readily written down: 
\FL
\begin{eqnarray}
\beta_u ( u, v, w ) = \beta_u ( u, v + 2w, -w ) \ \ , 
\qquad \qquad \qquad \ 
\nonumber \\
\beta_v ( u, v, w ) + 2\beta_w ( u, v, w ) = 
\beta_v ( u, v + 2w, -w ) \ \ , 
\label{eq:3.13} \\
\beta_w ( u, v, w ) = -\beta_w ( u, v + 2w, -w ) \ \ . 
\qquad \qquad \ \ \ \ 
\nonumber
\end{eqnarray}
and 
\FL
\begin{eqnarray}
\beta_u ( u, v, w ) + \beta_v ( u, v, w ) + 
2\beta_w ( u, v, w ) = \ \ \ \ 
\nonumber \\
= \beta_u ( u + v + 2w, -2w, -{v \over 2} ) \ \ , 
\nonumber \\
\beta_v ( u, v, w ) = -2\beta_w ( u + v + 2w, -2w, 
-{v \over 2} ) \ \ , 
\label{eq:3.14} \\
2\beta_w ( u, v, w ) = -\beta_v ( u + v + 2w, -2w, 
-{v \over 2} ) \ \ . 
\nonumber
\end{eqnarray}
One can see that expansions 
Eqs.~(\ref{eq:3.3a}, \ref{eq:3.3b}, \ref{eq:3.3c})
do really satisfy these relations. Moreover, due to this 
special symmetry, transformations Eqs.~(\ref{eq:3.10}) and 
(\ref{eq:3.12}) can, at most, rearrange the 
fixed points of RG equations not affecting 
numerical values of their coordinates $u_c$, $v_c$, and $w_c$ 
themselves. It provides powerful tool for evaluation of the 
accuracy of the approximation scheme employed. 

To calculate critical exponents field--theoretical expansions for 
two of them are needed. We find $\gamma^{-1}$ and $\eta$ as power 
series in $u$, $v$, and $w$ up to three--loop order. They are 
as follows:
\widetext
\FL
\begin{eqnarray}
\gamma^{-1} = 1 - {1 \over {N + 4}} \Bigl[ {{N + 1} \over 2} u + 
v + w \Bigr] + {1 \over {( N + 4 )^2}} 
\Bigl[ {{N + 1} \over 2} u^2 + 2 \bigl( uv + uw + 
vw \bigr) + v^2 + Nw^2 \Bigr] 
\nonumber \\
- {0.2472701 \over {( N + 4 )^3}} 
\Bigl[ \bigl( N^2 + 5 N + 4 \bigr) u^3 + 
\bigl( 6 N + 24 \bigr) \bigl( u^2v + u^2w + vw^2 \bigr) + 10 \bigl( 
3uv^2 + 6uvw \ \ \ 
\nonumber \\
+ v^3 + 3v^2w \bigr) 
+ \bigl( 18 N + 12 \bigr) uw^2 + \bigl( 2 N^2 + 8 \bigr) w^3 \Bigr] - 
{0.1925093 \over {( N + 4 )^3}} \Bigl[ \bigl( N^2 + 2 N + 1 \bigr) 
u^3 \ \ \ \ \ \ \nonumber \\
+ {18 ( N + 1 ) \over 3} \bigl( u^2v + u^2w \bigr) 
+ \bigl( 2 N + 10 \bigr) \bigl( uv^2 + 2uvw \bigr) + \bigl( 2 N^2 + 
2 N + 8 \bigr) uw^2 \qquad \ \ \ \ \ \ \ 
\nonumber \\
+ 4 \bigl( v^3 + 3v^2w + Nw^3 \bigr) + \bigl( 4 N + 8 \bigr) vw^2 
\Bigr] \ \ , 
\qquad \qquad \qquad \qquad \qquad \qquad \qquad \qquad \qquad \ 
\label{eq:3.15}
\end{eqnarray}
\FL
\begin{eqnarray}
\eta = {4 \over {27 ( N + 4 )^2}} \Bigl[ \bigl( N + 1 \bigr) u^2 + 
2 \bigl( 2uv + 2uw + v^2 + 2vw + Nw^2 \bigr) \Bigr] 
\qquad \qquad \qquad \qquad \qquad \ \ \ \ 
\nonumber \\
+ {0.01234194 \over {( N + 4 )^3}} \Bigl[ \bigl( N^2 + 5 N + 4 
\bigr) u^3 + \bigl( 6 N + 24 \bigr) \bigl( u^2v + u^2w + vw^2 \bigr) 
\qquad \qquad \qquad \qquad \ \ \ 
\nonumber \\
+ 10 \bigl( 3uv^2 + 6uvw + v^3 + 3v^2w \bigr) + 
\bigl( 18 N + 12 \bigr) uw^2 + \bigl( 2 N^2 + 8 \bigr) w^3 
\Bigr] \ \ . 
\qquad \qquad \qquad \ \ \ \ 
\label{eq:3.16}
\end{eqnarray}
\narrowtext
Since critical exponents are measurable (observable) quantities, the 
right--hand sides of Eqs.~(\ref{eq:3.15}) and (\ref{eq:3.16}) should 
contain for $N = 2$ only those combinations of $u$, $v$, and $w$ which 
are invariant under the transformations Eqs.~(\ref{eq:3.10}) and 
(\ref{eq:3.12}). As may be seen, it is actually the case.

\section{Numerical results}
\label{sec:4}

So, we perform the resummation of three--loop expansions 
Eqs.~(\ref{eq:3.3a},
\ref{eq:3.3b}, \ref{eq:3.3c}) by means of generalized 
Pad$\acute {\rm e}$--Borel 
technique with the approximant $[3/1]$ being used for the analytic 
continuation of Borel transforms. Coordinates of the fixed points 
of RG equations thus obtained are found numerically for the most 
interesting 
cases $N = 2$ and $N = 3$. They are presented in Table \ref{table1} 
($N = 2$) and Table \ref{table2} ($N = 3$) 
which contain also, for comparison, the fixed point 
coordinates obtained earlier \cite{55} from two--loop RG expansions 
resummed on the base of $[2/1]$ Pad$\acute {\rm e}$ approximants. 
Three--loop contributions are seen 
to change appreciably locations of all non--trivial fixed points.

Let us first discuss the numerical accuracy of the values found. 
The point~2 in Tables \ref{table1} and \ref{table2} 
is actually $O(2N)$--symmetric 
fixed point and its coordinates are to be compared with those 
obtained for $O(4)$-- and $O(6)$--symmetric models 
with real fields $\varphi_{\alpha}$
from resummed highest--order RG series available up today. Four--loop 
calculations in $3D$ have resulted in $u_c = 1.377$ for $n = 4$ and 
$u_c = 1.338$ for $n = 6$ \cite{64}. These numbers differ from their 
three--loop counterparts presented in second columns of 
Tables \ref{table1} and \ref{table2} by no more than 1\%. 
 
 Third columns of the Tables \ref{table1} and \ref{table2} contain 
coordinates of the Bose ($XY$) fixed point. The most accurate estimate
for $v_c$ obtained by the resummation of 
the six--loop 3D RG series is $v_c = 1.405$ \cite{1,2}.  
To compare this number with those presented in the Tables 
\ref{table1} and \ref{table2}, however, we should make some 
rescaling of $v_c$ for $N = 2$ and $N = 3$. 
The point is that the coefficient for $v^2$ in $\beta_v$ 
(Eq.~(\ref{eq:3.3b})) is equal to $5 \over{N + 4}$ 
differing from unity for $N \neq 1$. Since the
six--loop value of $v_c$ has been calculated in 
$O(2)$--symmetric model, i.e. for $N = 1$, 
the numbers in third columns of Table \ref{table1} and 
\ref{table2} should be multiplyed, before comparison, by factors 
$5 \over 6$ and $5 \over 7$ respectively. 
It gives $v_c = 1.4032$ ($N = 2$) and 
$v_c = 1.4033$ ($N = 3$). Practical coincidence of these two 
values is very 
natural since they are actually coordinates of the {\sl same} (Bose) 
fixed point while their closeness ($\sim 0.1\%$) to the six--loop 
value of $v_c$ provides an evidence of the high accuracy of the 
approximation scheme employed. 
Note, that the two--loop approximation leads to 
$v_c = 1.5583$ which is more than $10\%$ away from the ``exact'' value. 

Strong evidences of the high numerical accuracy of the approach 
elaborated may be obtained on the base of symmetry 
arguments. As was shown above, transformations 
Eqs.~(\ref{eq:3.10}) and (\ref{eq:3.12})
can only rearrange the fixed points of RG equations (\ref{eq:3.3a}, 
\ref{eq:3.3b}, \ref{eq:3.3c}) for $N = 2$ not affecting the values 
of $u_c$, $v_c$, and $w_c$ themselves. Indeed, this is precisely 
what occurs when one applies Eq.~(\ref{eq:3.10}) 
to the content of Table \ref{table1}: 
the first four fixed points stay at their places while the 
points 5--8 undergo pair transpositions 
$5 \to 7$, $7 \to 5$, $6 \to 8$, $8 \to 6$. 
Another transformation, Eq.~(\ref{eq:3.12}), practically does not 
change the location of fixed 
points 1, 2, 7, and 8 and causes pair transposition $3 \to 5$, 
$5 \to 3$. 
The rest of fixed points, the 4--th and the 6--th ones, however, are 
converted one to another under Eq.~(\ref{eq:3.12}) only within 
three--loop 
approximation. Corresponding two--loop results turn out to violate the 
symmetry relations induced by Eq.~(\ref{eq:3.12}). More precisely, the 
differences between the coordinates of the point 4 and the transformed 
coordinates of the point 6 (``symmetry discrepancies'') 
given by $[2/1]$ 
Pad$\acute {\rm e}$--Borel approximants are about 0.3, while within 
three--loop approximation they are of order of 0.01. 

So, the three--loop terms being taken into account enable one to 
obtain 
results which are much more accurate than those given by two--loop RG 
expansions. Moreover, it is seen that the field--theoretical RG 
approach in three dimensions combined with a generalized 
Pad$\acute {\rm e}$--Borel resummation technique does really 
provide a 
regular, rapidly converging approximation scheme powerful enough to 
treat complicated model with three quartic coupling constants. 
At the same time, the Chisholm--Borel resummation 
procedure is found to give poor results in this case.

Let us discuss further the stability of the fixed points and the 
structure 
of the RG flow diagrams. All fixed points of the RG equations are 
unstable in the three--dimensional parameter space ($u, v, w$). 
The 4--th and the 6--th ones, however, 
are stable within the planes ($u, v$) and ($u, w$) 
respectively. The existence of such points is important since it 
implies 
the possibility of continuous phase transitions in numerous physical 
systems described by the model Eq.~(\ref{eq:1.1}) with $N = 2$ and 
$v_0 = 0$ or $w_0 = 0$. 
RG flows for $N = 2$ within the planes ($u, v$) and ($u, w$) 
and for $N = 3$ within the plane ($u, v$) are shown in 
Fig.~\ref{fig1}(b) and Fig.~\ref{fig2}(b, c). 
One can see from these figures that there is 
not a fixed point stable within the plane ($u, w$) for $N = 3$ 
while for 
$N = 2$ such a point exists. Hence, the topology of the flow diagram 
should change when $N$ varies. It is interesting, therefore, to study 
the structure of our RG flows for arbitrary $N$.

The detailed numerical analysis of three--loop RG equations obtained 
shows that only two diverse $u-v$ flow pictures 
occur for $1 < N < \infty$ while the RG 
flow within the plane ($u, w$) may proceed in four different ways. 
All possible scenarios are depicted in Fig.~\ref{fig1} 
and Fig.~\ref{fig2} (see hardcopy: Phys. Rev. B {\bf 49}, 15901 (1994)). 
The critical (marginal) dimensionality of the 
order parameter $N_c$ which separates from each other 
two regimes of RG flows for $w = 0$ is found to be: 
\begin{equation}
N_c = 1.47 \pm 0.01 \ \ . \label{eq:4.1} 
\end{equation}
Since this number is less than two, in all physically interesting 
cases, 
i.e. for $N \ge 2$, the $O(2N)$--symmetric fixed point turns out 
to be 
unstable. So, the system should undergo either a continuous phase 
transition demonstrating an anisotropic ($v_c \neq 0$) 
critical behavior or a fluctuation 
induced first--order phase transition. When $N \to \infty$ the 
anisotropic stable fixed point in the plane ($u, v$) 
is going to the $O(2)$--symmetric 
one which becomes degenerate in this spherical--model limit. 

The behavior of our model in the plane ($u, w$) is more rich. It is 
characterized by three marginal values of the order parameter 
dimensionality: 
$N_{c1}$, $N_{c2}$, and $N_{c3}$. Calculated on the base of resummed 
three--loop RG series Eqs.~(\ref{eq:3.3a}, \ref{eq:3.3b}, 
\ref{eq:3.3c}) they are as follows: 
\begin{eqnarray}
N_{c1} = 1.45 \pm 0.01 \ \ , \nonumber \\
N_{c2} = 2.03 \pm 0.01 \ \ , \label{eq:4.2} \\
N_{c3} = 3.91 \pm 0.01 \ \ . \nonumber
\end{eqnarray}
For $N < N_{c1}$ (Fig.~\ref{fig2}(a)) three non--trivial fixed points 
exist in the plane ($u, w$) with the $O(2N)$--symmetric point being 
stable. When $N$ exceeds $N_{c1}$ this ``Heisenberg'' fixed point 
loses 
its stability but the other, anisotropic fixed point with $w_c < 0$ 
acquires it (Fig.~\ref{fig2}(b)). In this domain 
which includes the important case $N = 2$, our system demonstrates an 
anisotropic scaling behavior 
or discontinuous phase transitions. With increasing $N$ the stable 
fixed point in Fig.~\ref{fig2}(b) is moving downward and 
``annihilates'' 
with the anisotropic saddle fixed point when $N$ approaches $N_{c2}$. 
There is only one non--trivial fixed point in the domain 
$N_{c2} < N < N_{c3}$ including $N = 3$ (Fig.~\ref{fig2}(c)); it is 
$O(2N)$--symmetric and unstable. So, only first--order phase 
transitions are possible, in principle, 
in this case provided $w_0 \neq 0$. At last, 
when $N$ increases further and crosses over the value $N_{c3}$ the 
creation of two new anisotropic fixed points in the $u-w$ flow diagram 
takes place (Fig.~\ref{fig2}(d)). One of them is stable and describes 
some anisotropic critical behavior with $w_c > 0$. This fixed point is 
known as a ``chiral'' point \cite{12} and corresponding ``chiral'' 
phase 
transition has been extensively studied during the last years. 
As follows from our estimates (Eq.~(\ref{eq:4.2})), this point does 
really exist and governs the scaling behavior of physical systems 
with $N \ge 4$. 
For $N = 2$ and $N = 3$ the chiral critical behavior does not actually 
realized. 

Let's discuss numerical estimates Eqs.~(\ref{eq:4.1}) and 
(\ref{eq:4.2}) in more detail. 
The value of $N_{c2}$ turns out to be very close to $N = 2$ 
which is of prime physical importance. Can higher--order contributions 
to the $\beta$--functions being taken into account change $N_{c2}$, 
invert 
the inequality $N_{c2} > 2$, and, hence, alter the structure of the 
$u-w$ flow diagram for $N = 2$? {\sl No, they can not}. 
The point is that the structures of the RG flows 
in the planes ($u, v$) and ($u, w$) are related to each other 
for $N = 2$ by the symmetry relations discussed earlier. In 
particular, as may be seen from Eq.~(\ref{eq:3.12}) the total number 
of fixed points in each of these flow diagrams should be just the 
same. Since the plane ($u, v$) definitely contains four fixed points 
($N_c$ lies far below the value of interest $N = 2$) 
the plane ($u, w$) for $N = 2$ should possess 
four fixed points too. Moreover, since, for $N =2$, 
the stable fixed point 
has $v_c > 0$ its counterpart in the plane ($u, w$) should possess 
$w_c < 0$ (see Eq.~(\ref{eq:3.12})).
It means that inevitably $N_{c2} > 2$ in the exact theory. The 
estimate Eq.~(\ref{eq:4.2}) is in accord with this inequality. 

Another point to be discussed is the near coincidence of the 
calculated 
values of $N_c$ and $N_{c1}$. It is not occasional. Indeed, 
$N_c$ and $N_{c1}$ are both the values of $N$ for which the 
$O(2N)$--symmetric fixed point becomes degenerate and critical 
exponents 
describing its stability change a sign. These exponents are 
completely determined by derivatives 
$\partial \beta_v \over {\partial v}$ 
and $\partial \beta_w \over {\partial w}$ taken at the ``Heisenberg'' 
fixed point since $\partial \beta_v \over {\partial w}$ and 
$\partial \beta_w \over {\partial v}$ at this point vanish. 
One can see from Eqs.~(\ref{eq:3.3b}) and (\ref{eq:3.3c}), however, 
that 
${\partial \beta_v \over {\partial v}} = {\partial \beta_w \over 
{\partial w}}$ along the whole line $v = w = 0$ up to the highest 
calculated order. So, when $N$ varies the ``Heisenberg'' fixed point 
should 
lose its stability in planes ($u, v$) and ($u, w$) simultaneously, 
i.e. $N_c$ and $N_{c1}$ should be equal to each other. The small 
difference between calculated values of 
$N_c$ and $N_{c1}$ reflects a finite accuracy 
of our approximation scheme which is seen to be of order of 0.01. 

Having calculated fixed point coordinates we can find critical 
exponents for our model. To obtain accurate estimates for 
$\gamma$ the expansion 
Eq.~(\ref{eq:3.15}) is resummed by means of the generalized 
Pad$\acute {\rm e}$--Borel procedure described above while the values 
of $\eta$ are found by direct substitution of fixed point coordinates 
into 
Eq.~(\ref{eq:3.16}) since this very short series with very small and 
positive three--loop term doesn't need in resummation. The results 
obtained 
for $N = 2$ and $N = 3$ are presented in Table \ref{table3} and Table 
\ref{table4} respectively which contain also the values of $\gamma$ 
and $\eta$ calculated earlier \cite{55} within two--loop 
approximation. 

Three--loop contributions are seen to change critical exponents 
values only slightly. For $N = 2$ critical exponents 
calculated in the fixed points 
3, 5, and 7 turn out to be almost identical, and so is true for the 
fixed points 4, 6, and 8. In the exact theory each of these two sets 
of fixed points indeed should possess identical critical exponents 
since 
fixed points belonging to the same set are related to each other by 
symmetry relations Eqs.~(\ref{eq:3.10}) and (\ref{eq:3.12}), i.e. 
they are 
actually the same fixed point. So, differences between the values of 
$\gamma$ and $\eta$ calculated in such fixed points may be 
considered as a measure of numerical accuracy of our approximation. 
It is seen to be better than 0.001. On the other hand, the difference 
between the values of $\gamma$ 
calculated in the fixed points 4 and 6 (or 8) within two--loop 
approximation exceeds 0.05. Hence, the taking into account of 
three--loop terms improves the situation essentially. 

It is worthy also to 
compare critical exponents found in the Bose and ``Heisenberg'' fixed 
points with 
their counterparts determined from six--loop \cite{1,2} and four--loop 
\cite{64} RG expansions for a simple $O(n)$--symmetric model. The most 
accurate estimate for the susceptibility exponent of the $3D$ $XY$ 
model is $\gamma = 1.315$ \cite{1,2}.
Corresponding values in Tables \ref{table3} and \ref{table4} 
(third columns) differ from it by 0.005. Four--loop RG calculations 
for $n = 2N = 4$ and $n = 2N = 6$ give 
$\gamma = 1.441$ and $\gamma = 1.541$ respectively \cite{64}.
Differences between these numbers and their three--loop twins 
presented in Tables \ref{table3} and \ref{table4} (second columns) 
are about 0.02. So, we arrive to the conclusion: 
Pad$\acute {\rm e}$--Borel resummed $3D$ 
three--loop RG expansions provide an accuracy of order of 0.01 for 
all calculated quantities. This accuracy is sufficient 
for making definite and reliable theoretical prediction for 
physical systems described by the model 
Eq.~(\ref{eq:1.1}). It will be done in the following Section. 

Now let's return back to the calculation of critical exponents. 
The rest of them may be found by making use of well-known 
scaling relations. We present here numerical values of exponents 
$\nu$, $\alpha$, and $\beta$ for the fixed points which are stable
within corresponding parameter subspaces since only these numbers
may be related to experiments. So, for equivalent fixed points 
4, 6, and 8 from Table \ref{table3}  
\begin{equation}
\nu = 0.677 \ \ , \quad \alpha = -0.030 \ \ , \quad \beta = 0.347 
\label{eq:4.3}
\end{equation}
while the point 4 in Table \ref{table4} is  
characterized by $\nu = 0.673$, $\alpha = -0.020$, 
and $\beta = 0.345$.

\section{Application to physical systems and discussion}
\label{sec:5}

All fixed points of our RG equations were found to be unstable within 
the three--dimensional parameter space ($u, v, w$) for $N = 2$ and 
$N = 3$. It means that only discontinuous, first--order phase 
transitions should occur in physical systems with 
non--zero initial values of $v$ and $w$. 
Such systems are represented by cubic and tetragonal unconventional 
superconductors and superconductors with composite $s - d$ order 
parameters. On the other hand, fluctuation--driven first--order 
phase transitions are 
known to be extremely weak. So, the absence, within the experimental 
accuracy, of discontinuous superconducting transitions in relevant 
heavy--fermion and high--$T_c$ compounds does not actually contradict 
to the above conclusion. 

In hexagonal $d$--wave superconductors described by the model 
Eq.~(\ref{eq:1.1}) with $N = 2$ and $v_0 = 0$ second--order phase 
transitions remain possible under strong superconducting 
fluctuations since there is 
a stable fixed point within the plane ($u, w$) which possesses 
a sizable domain of attraction. Corresponding values of critical 
exponents (column 4 (6, 8) in Table \ref{table3} and 
Eq.~(\ref{eq:4.3})) turn out to be close enough to those 
of the $3D$ $XY$ model. So, it is actually impossible to 
distinguish between the BCS $s$--wave pairing and the non--trivial one 
studying experimentally the scaling behavior of superconductors. On 
the other hand, anisotropic gradient terms omitted in the Hamiltonian 
(\ref{eq:1.1}) can themselves change, in course of fluctuation 
renormalization, the order of phase transition and the structure of 
phase diagram, as they do in crystals undergoing structural 
(ferroelectric) and magnetic phase transitions 
\cite{65,66}. This will obviously result in a 
non--universal behavior of hexagonal $d$--wave superconductors in the 
critical region. 

In liquid helium--3, where $N = 3$ and $v_0 = 0$ and, therefore, RG 
equations have no stable fixed points, fluctuations should always 
force 
the superfluid phase transition to be first--order. Corresponding 
discontinuities of thermodynamic quantities at the transition point, 
however, would hardly be observed experimentally because of the 
narrowness of the critical region in this Fermi--liquid 
(see, e.g. Refs. \cite{16,67} for numerical estimates). 

Only first--order phase transitions should emerge also in FFJJA at 
$T = 0$, in spite of the existence of stable fixed points in planes 
($u, v$) and 
($u, w$) for $N = 2$. Indeed, RG trajectories starting from physical 
initial points, i.e. from those having $v_0 < 0$ and $w_0 > 0$ for 
triangular and square FFJJA respectively \cite{10,60a}, can not achieve 
the stable fixed points as is cleary seen from Fig.~\ref{fig1}(b) and 
Fig.~\ref{fig2}(b). So, these systems will demonstrate non--universal 
critical behavior. 

A mode of the critical behavior of STA and HM described by the 
Hamiltonian (\ref{eq:1.1}) with $v_0 = 0$ depends on the 
dimensionality of the order 
parameter. In materials with Heisenberg spins, i.e. for $N = 3$, only 
(weak) first--order phase transitions should occur. In easy--plane 
crystals 
with $XY$--like spins continuous transitions are also possible with 
critical exponents presented in column 4 (6, 8) of Table \ref{table3} 
and 
Eq.~(\ref{eq:4.3}) which are practically undistinquishable from those 
of $3D$ $XY$ model. These exponents, however, govern transitions into 
somewhat trivial phases: simple ferromagnetic or antiferromagnetic in 
STA and a sinusoidal (linearly--polarized) in HM, since relevant 
stable 
fixed point possesses $w_c < 0$. Much more interesting ordering with 
frustration in STA and helical one in HM are described by 
Eq.~(\ref{eq:1.1}) 
with $w_0 > 0$. They may be realized only via first--order phase 
transitions, as is clearly seen from Fig.~\ref{fig2}(b). 

We did not find any traces of chiral second--order phase 
transitions and corresponding new classes of universality for $N = 2$ 
and $N = 3$, i.e. for 
STA and HM with Heisenberg or $XY$--like spins. This result is in 
contradiction with conjectures and conclusions made on the base of the 
lower--order $\epsilon$--expansion analysis \cite{12,58}. Such 
conclusions, 
however, can not be referred to as reliable since the method mentioned 
provides rather low numerical accuracy in three dimensions. 
To illustrate this point and to clear up the situation let's discuss 
two--loop $\epsilon$--expansions (highest--order now available) for 
marginal order--parameter dimensionalities $N_{c1}$, $N_{c2}$, and 
$N_{c3}$. They are as follows \cite{12}:
\begin{mathletters}
\begin{equation}
N_{c1} = 2 - \epsilon \ \ , \label{eq:5.1a} \qquad \ \ \ 
\end{equation}
\begin{equation}
N_{c2} = 2.20 - 0.57 \epsilon \ \ , \label{eq:5.1b} 
\end{equation}
\begin{equation}
N_{c3} = 21.8 - 23.4 \epsilon \ \ . \label{eq:5.1c}
\end{equation}
\end{mathletters}
When $\epsilon \to 1$ $N_{c3}$ becomes less than 2 and chiral fixed 
point seems to exist for $N = 2$ and $N = 3$. In this limit, 
however, $N_{c2}$ becomes less than 2 too what is in obvious 
contradiction with an inequality $N_{c2} > 2$ 
proven above. Moreover, another inequality $N_{c2} < N_{c3}$ 
valid for $\epsilon \ll 1$ turns out to be broken at $\epsilon = 1$ 
as well. 

Is it possible to make $\epsilon$--expansion predictions more accurate 
for $\epsilon = 1$? Yes, of course. Higher--order (four-- and 
five--loop) 
$\epsilon$--expansions are known to give rather good numerical results 
for $3D$ $O(n)$--symmetric model at $\epsilon = 1$ provided some 
Borel--like resummation procedure is applied \cite{69,70,71}. 
Unfortunately, 
we have no long enough $\epsilon$--expansions for our model up today 
\cite{72}. So, we try to ``sum up'' expansions (\ref{eq:5.1a}, 
\ref{eq:5.1b}, 
\ref{eq:5.1c}) constructing simple Pad$\acute {\rm e}$ approximants: 
\begin{mathletters}
\begin{equation}
N_{c1} \approx {2 \over {1 + 0.5 \epsilon}} \ \ , \label{eq:5.2a}
\end{equation}
\begin{equation}
N_{c2} \approx {2.20 \over {1 + 0.26 \epsilon}} \ \ , \label{eq:5.2b}
\end{equation}
\begin{equation}
N_{c3} \approx {21.8 \over {1 + 1.07 \epsilon}} \ \ . \label{eq:5.2c}
\end{equation}
\end{mathletters}
For $\epsilon = 1$ these formulas give $N_{c1} = 1.33$, 
$N_{c2} = 1.75$, 
and $N_{c3} = 10.5$. The first number is much closer to our estimate 
$N_{c1} = 1.45$ (Eq.~(\ref{eq:4.2})) than the value $N_{c1} = 1$ 
given by Eq.~(\ref{eq:5.1a}). The second one is also closer to the 
three--loop $3D$ estimate 
$N_{c2} = 2.03$ than the naive value $N_{c2} = 1.63$, but both 
violate the inequality $N_{c2} > 2$. The third number exceeds 
enormously the estimate $N_{c3} = 3.91$ which turns out to lie 
between this number 
and the naive estimate $N_{c3} = -1.6$. So, we see that primitive 
resummation of very short expansions (\ref{eq:5.1a}, \ref{eq:5.1b}, 
\ref{eq:5.1c}) results in somewhat improved numerical estimates for 
$N_{c1}$ and $N_{c2}$ while being used for evaluation of $N_{c3}$ it 
demonstrates that lower--order $\epsilon$--expansions are useless in 
this case. Hence, lower--order calculations in $4 - \epsilon$ 
dimensions can not 
be considered as a tool for answering the question whether $3D$ 
physical systems with $N = 2$ and $N = 3$ undergo chiral phase 
transitions or not. 

Monte--Carlo simulations \cite{11,28} would also hardly be referred 
to as 
evidence of chiral critical behavior of STA and HM with Heisenberg or 
$XY$--like spins. The point is that unusual values of critical 
exponents 
given by such calculations turn out to be close to tricritical ones. 
That's why it was suggested \cite{29} that tricritical behavior or 
tricritical--to--critical crossover are really seen in these computer 
experiments as well as in most of physical experiments performed on 
several 
helimagnets (Tb, Dy, Ho) and STA (CsMnBr$_3$, CsVCl$_3$ and others). 
We completely agree with what is argued on this topic in 
Ref.~\cite{29} 
where the reader can find also an overview and analysis of relevant
experimental data. 

\section{Conclusions}
\label{sec:6}

The critical behavior of the model describing phase transitions 
in superconducting and magnetic systems with complex $N$--vector 
order parameter as well as in superfluid helium--3 has been 
studied within the RG approach in three dimensions. RG 
$\beta$--functions and critical exponents have been calculated 
as series in powers of renormalized quartic coupling constants 
$u$, $v$, and $w$ up to three--loop order. The series obtained 
have been resummed by means of the generalized 
Pad$\acute {\rm e}$--Borel technique and fixed points coordinates, 
critical exponents values and a structure of RG flows have been 
determined for $N = 2$ and $N = 3$. Marginal values of the order 
parameter dimensionality at which the topology of RG flows in 
planes ($u, v$) and ($u, w$) changes have been also found. Several 
criterions have been used to estimate the accuracy of numerical 
results obtained which had turned out to be about 0.01, an order 
of magnitude better than that given by resummed two--loop RG 
expansions. So, field--theoretical RG approach in three dimensions 
combined with a proper resummation technique provides a regular, 
rapidly converging approximation scheme powerful enough to treat 
complicated model with three quartic coupling constants. 

Relevant physical systems have been shown to undergo, in most 
cases, fluctuation induced first--order phase transitions. 
Second--order transitions have been found to occur only in 
hexagonal $d$--wave superconductors and in planar magnets 
(into somewhat trivial phases: linearly--polarized or 
unfrustrated). Corresponding critical exponents have turned out 
to differ from those of the $3D$ $XY$ model by no more than 
0.02 -- 0.03, i.e. the underlying critical behavior would hardly 
be thought as experimentally distinguishable from the Bose one. 
RG equations obtained have been shown to possess the chiral fixed 
point but only for $N \ge 4$. It means that STA and HM with 
Heisenberg and $XY$--like spins would not really demonstrate 
the chiral critical behavior with unusual critical exponents 
approaching helical or frustrated antiferromagnetic states via 
first--order phase transitions. 

\acknowledgements

We gratefully acknowledge useful discussions with B.~N.~Shalayev. 
One of us 
(S.~A.~A.) would also thank D.~A.~Zuev for providing a considerable 
amount of computer time. This work was supported in part by the 
Scientific Council on the Problem of High Temperature 
Superconductivity 
via the Russian National Program ``High Temperature 
Superconductivity'' 
under Project No.~90476 ``Gran'', and in part by the Science and 
Higher 
School State Committee of Russian Federation through Grant 
No.~93--7.1--51. 

\unletteredappendix{}

In this Appendix, some details of the resummation procedure employed 
are described. As was shown in Sec.~\ref{sec:3}, resolvent series 
\begin{eqnarray}
\tilde F (x, y, z; \lambda) = \sum\limits_{n = 0}^{\infty} A_n 
\lambda^n \ \ ,
\qquad \ \ \ \ \ \nonumber \\ \label{eq:a1} \\
A_n = \sum\limits_{l = 0}^n \sum\limits_{m = 0}^{n - l} 
{c_{l, m, n - l - m} \over{n !}} x^l y^m z^{n - l - m} 
\nonumber
\end{eqnarray}
for Borel transforms of the original multi--variable RG expansions 
may be constructed to generate Pad$\acute {\rm e}$ approximants 
$[L/M]$ in parameter $\lambda$. 
These approximants are defined in a conventional way:
\begin{equation}
[L/M] = {P_L (\lambda) \over {Q_M (\lambda)}} \ \ , \label{eq:a2}
\end{equation}
where $P_L (\lambda)$ and $Q_M (\lambda)$ are polynomials of 
degrees $L$ 
and $M$ respectively with coefficients depending on $x$, $y$, and $z$, 
which may be determined from the following relations: 
\begin{eqnarray}
Q_M (\lambda) \tilde F (x, y, z; \lambda) - P_L (\lambda) = 
O (\lambda^{L + M + 1}) \ \ , \nonumber \\ \label{eq:a3} \\
Q_M (0) = 1 \ \ . 
\qquad \qquad \qquad \qquad \qquad \qquad \ \ \ \ \ 
\nonumber
\end{eqnarray}
Approximate expressions for $\beta$--functions and critical exponents 
are 
then obtained by the replacement of variables $x = ut$, $y = vt$, and
$z = wt$ in Pad$\acute {\rm e}$ approximants and by evaluation of the 
integral 
\begin{equation}
\int\limits_0^{\infty} e^{-t} [L/M] \Big \arrowvert_{\lambda = 1} dt 
\label{eq:a4}
\end{equation}
(Borel transformation). 

With three--loop expansions in hand, we can use two different 
approximants 
$[3/1]$ and $[2/2]$ obeying the condition $L \ge M$. The former was 
shown (Sec.~\ref{sec:4}) to provide rather good numerical results for 
all cases considered. Moreover, an employment of this approximant 
kept us away from the well--known problem 
of poles which often arises when approximants 
with higher--order denominators are used. That's why we have chosen 
Pad$\acute {\rm e}$ approximant $[3/1]$ for our analysis. When 
expressed 
in terms of renormalized coupling constants $u$, $v$, and $w$ and the 
variable $t$ it is as follows: 
\begin{equation}
[3/1] = {a_0 + a_1 t + a_2 t^2 + a_3 t^3 \over {1 + b_1 t}} \ \ , 
\label{eq:a5}
\end{equation}
where $a_0, \ldots, a_3$ and $b_1$ are known functions of $u$, 
$v$, and $w$. If series to be resummed are those for 
$\beta$--functions the 
coefficient $a_0$ in Eq.~(\ref{eq:a5}) turns out to vanish and the 
integral (\ref{eq:a4}) reads 
\begin{equation}
\int\limits_0^{\infty} t e^{-t} {a_1 + a_2 t + a_3 t^2 
\over {1 + b_1 t}} dt \ \ . 
\label{eq:a6}
\end{equation}
Evaluating this integral we get the final expression (the ``sum'' 
of the series) for the function of interest: 
\FL
\begin{eqnarray}
f (u, v, w) = (a_1 + a_2 + 2a_3) b - (a_2 + a_3 - a_3b) b^2 
\nonumber \\
+ (a_1 - a_2b + a_3b^2) b^2 e^b Ei(-b) \ \ , \ \ \ \ \ \label{eq:a7}
\end{eqnarray}
where $Ei(x)$ is the exponential integral \cite{100} and 
$b = b_1^{-1}$.
This is precisely the formula which was used for resummation of the 
three--loop RG expansions Eqs.~(\ref{eq:3.3a}, \ref{eq:3.3b},  
\ref{eq:3.3c}) and for determination of the fixed points. 
The approximate expression for $\gamma^{-1}(u, v, w)$ is quite 
similar and not presented here.

\figure{RG flows in the plane ($u, v$) for $N < N_c$ and $N > N_c$ 
where $N_c = 1.47 \pm 0.01$. Shaded areas represent the regions 
of instability of the Hamiltonian (\ref{eq:1.1}) (see hardcopy: 
Phys. Rev. B {\bf 49}, 15901 (1994)).
\label{fig1}}

\figure{Four possible scenarios of RG flow in the plane ($u, w$). 
Marginal values of the order parameter dimensionality $N_{c1}$, 
$N_{c2}$, and $N_{c3}$ are given by Eq.~(\ref{eq:4.2}). Shaded 
areas are the regions of instability of the Hamiltonian 
(\ref{eq:1.1}) (see hardcopy: Phys. Rev. B {\bf 49}, 15901 (1994)). 
\label{fig2}}

\widetext
\begin{table}
\caption{Coordinates of the fixed points of RG equations for $N = 2$ 
obtained within three--loop (approximant $[3/1]$) and two--loop 
(approximant $[2/1]$) approximations.}
\begin{tabular}{cccccccccc}
 & &1&2&3&4&5&6&7&8 \\
\tableline
$u_c$ & $[3/1]$ & 0.0 & 1.3671 & 0.0 & 0.1872 & 1.6833 & 1.6787 
& 1.6832 & 1.6789 \\
& $[2/1]^{*}$ & 0.0 & 1.4863 & 0.0 & 0.0340 & 1.8699 & 1.8334 
& 1.8699 & 1.8334 \\ 
\tableline
$v_c$ & $[3/1]$ & 0.0 & 0.0 & 1.6838 & 1.4914 & 0.0 & 0.0 & -1.6800 
& -1.4950 \\
& $[2/1]^{*}$ & 0.0 & 0.0 & 1.8699 & 1.8334 & 0.0 & 0.0 & -1.8699 
& -1.3591 \\
\tableline
$w_c$ & $[3/1]$ & 0.0 & 0.0 & 0.0 & 0.0 & -0.8416 & -0.7477 & 0.8400 
& 0.7480 \\
& $[2/1]^{*}$ & 0.0 & 0.0 & 0.0 & 0.0 & -0.9350 & -0.6796 & 0.9349 
& 0.6795 \\
\end{tabular}
\label{table1}
\tablenotes{$^{*}$ Quoted from Ref.~\cite{55}}
\end{table}

\narrowtext
\begin{table}
\caption{Coordinates of the fixed points of RG equations for $N = 3$ 
obtained within three--loop (approximant $[3/1]$) and two--loop 
(approximant $[2/1]$) approximations.}
\begin{tabular}{cccccc}
& & 1 & 2 & 3 & 4 \\
\tableline
$u_c$ & $[3/1]$ & 0.0 & 1.3310 & 0.0 & 0.0780 \\
& $[2/1]^{*}$ & 0.0 & 1.4262 & 0.0 & 0.0097 \\
\tableline
$v_c$ & $[3/1]$ & 0.0 & 0.0 & 1.9646 & 1.8845 \\
& $[2/1]^{*}$ & 0.0 & 0.0 & 2.1816 & 2.1713 \\
\tableline
$w_c$ & $[3/1]$ & 0.0 & 0.0 & 0.0 & 0.0 \\
& $[2/1]^{*}$ & 0.0 & 0.0 & 0.0 & 0.0 \\
\end{tabular}
\label{table2}
\tablenotes{$^{*}$ Quoted from Ref.~\cite{55}}
\end{table}

\widetext
\begin{table}
\caption{Critical exponents $\gamma$ and $\eta$ for $N = 2$ calculated 
within three--loop (approximant $[3/1]$) and two--loop (approximant 
$[2/1]$) approximations.}
\begin{tabular}{cccccccccc}
& & 1 & 2 & 3 & 4 & 5 & 6 & 7 & 8 \\
\tableline
$\gamma$ & $[3/1]$ & 1 & 1.4260 & 1.3099 & 1.3360 & 1.3098 & 1.3355 & 
1.3102 & 1.3357 \\
& $[2/1]^{*}$ & 1 & 1.4347 & 1.3218 & 1.3259 & 1.3218 & 1.3799 
& 1.3218 & 1.3799 \\
\tableline
$\eta$ & $[3/1]$ & 0 & 0.0257 & 0.0261 & 0.0261 & 0.0260 & 0.0261 
& 0.0260 & 0.0261 \\
& $[2/1]^{*}$ & 0 & 0.0273 & 0.0288 & 0.0287 & 0.0288 & 0.0286 
& 0.0288 & 0.0286 \\
\end{tabular}
\label{table3}
\tablenotes{$^{*}$ Quoted from Ref.~\cite{55}}
\end{table}

\narrowtext
\begin{table}
\caption{Critical exponents $\gamma$ and $\eta$ for $N = 3$ 
calculated 
within three--loop (approximant $[3/1]$) and two--loop (approximant 
$[2/1]$) approximations.}
\begin{tabular}{cccccc}
& & 1 & 2 & 3 & 4 \\
\tableline
$\gamma$ & $[3/1]$ & 1 & 1.5164 & 1.3099 & 1.3291 \\
& $[2/1]^{*}$ & 1 & 1.5217 & 1.3218 & 1.3220 \\
\tableline
$\eta$ & $[3/1]$ & 0 & 0.0238 & 0.0261 & 0.0261 \\
& $[2/1]^{*}$ & 0 & 0.0246 & 0.0288 & 0.0286 \\
\end{tabular}
\label{table4}
\tablenotes{$^{*}$ Quoted from Ref.~\cite{55}}
\end{table}


\begin{references}
\bibitem {43b} We mean here high--temperature superconductors with
moderate anisotropy such as YBa$_2$Cu$_3$O$_7$ which demonstrate 
well--pronounced 3D behavior in the vicinity of T$_c$.
\bibitem {72} Three--loop calculations for the model 
Eq.~(\ref{eq:1.1}) 
in $4-\epsilon$ dimensions are now in progress.
\bibitem {100} The following definition of $Ei (x)$ is adopted here: 
$Ei (x) = \int \limits_{- \infty}^x e^t t^{-1} dt$.
\end{references}
\end{document}